\documentstyle[12pt]{article}
\begin{document}
\baselineskip=15pt
\parskip=6pt
\title{Stability of the vortex lattice in a rotating superfluid}
\author{
Gordon Baym\\
Loomis Laboratory of Physics\\
University of Illinois at Urbana-Champaign\\
1110 W. Green St.\\
Urbana, IL 61801, U.S.A.\\}
\maketitle

\baselineskip=12pt
We analyze the stability of the vortex lattice in a rotating
superfluid against thermal fluctuations associated with the long-wavelength
Tkachenko modes of the lattice.  Inclusion of only the two-dimensional modes
leads formally to instability in infinite lattices; however, when the full
three-dimensional spectrum of modes is taken into account, the
thermally-induced lattice displacements are indeed finite.


\section{Introduction}

\baselineskip=15pt
\parskip=6pt

    A rotating $^4$He superfluid accomodates angular momentum by forming an
array of vortex lines, each having quantized circulation $2\pi\hbar/m$, where
$m$ is the $^4$He mass.  In a classic series of papers \cite{Tk}, Tkachenko
showed that the lowest energy configuration of an infinite vortex array is a
two-dimensional triangular lattice, and that the lattice supports collective
elastic modes; perpendicular to the rotation axis the modes travel with
velocity $(\hbar\Omega/4m)^{1/2}$, where $\Omega$ is the rotational velocity
of the fluid.  Finite arrays, as Campbell and Ziff predicted \cite{CZ}, should
exhibit distortions from triangular, structure seen experimentally by Packard
and co-workers \cite{Pa}. The question we address in this paper is whether the
infinite vortex lattice remains stable in the presence of thermal fluctuations
associated with Tkachenko shear modes.

    As in a two-dimensional system, where long-wavelength modes forbid strict
long-range translational order \cite{Me}, the long-wavelength Tkachenko
motions of the lattice, when limited to the plane perpendicular to the
rotation axis, lead to a weak logarithmic divergence of the displacements
characteristic of a two-dimensional solid.  However, when one includes the
full three-dimensional spectrum of excitations of the vortex lattice, with
line-bending contributions to the Tkachenko modes [5-8], stability of the
lattice is restored.  The vortex lattice provides an instructive realization
of how addition of a third dimension leads to stability of the ordering even
if the order parameter varies in only two dimensions; the results are
consistent with the usual Landau-Peierls stability arguments \cite{LP} in a
three-dimensional system, which forbid an order parameter varying in only one
dimension, but do not rule out stability for the two-dimensional variations
encountered in the superfluid vortex lattice.  Compared to a normal lattice,
however, the vortex system has the peculiarity that the effective energy
associated with lattice displacements depends on the displacements directly,
rather than simply on their derivatives; this dependence, arising, as seen
below, from the intrinsic lack of Galilean invariance of a rotating system,
acts to reduce the contribution of long-wavelength excitations to the
excursions of the vortices from their equilibrium sites.  Understanding how
stability is achieved in the vortex lattice is also instructive for the
closely related issue of whether thermally-excited elastic shear modes drive
instability of the Abrikosov magnetic-flux lattice in clean Type II
superconductors and in the layered high T$_c$ materials in particular
\cite{Moore}.  Extension of the present analysis to flux lattices will be
given in a subsequent paper \cite{typeII}.

    To study the stability we apply the familiar methodology \cite{Me,LP,Ho}
of calculating the contributions of the thermal fluctuations to the
displacements of the vortices from their equilibrium sites by constructing the
displacement autocorrelation functions in terms of the spectrum of modes
\cite{melt}.  In the following section we review the equations of motion of
the vortex lattice; in Sec. 3 we derive the displacement autocorrelation
functions of the lattice, and discuss the consequences for lattice stability
and phase coherence.  Appendix A derives the relation between the
macroscopically averaged phase of the order parameter, flow velocity and
vortex lattice displacements, and in Appendix B we analyze how the
non-vanishing commutation relations obeyed by the components of the lattice
displacements arise as one takes the limit in which the inertial mass of the
vortex lines vanishes.

\section{Equations of motion and long-wavelength modes}

    We consider an incompressible, non-dissipative superfluid in a bucket
rotating about the $z$-axis at angular velocity $\vec \Omega$, and work in the
frame rotating with the bucket; we follow the formulation given in Refs.
\cite{BC1} and \cite{BC2} of the complete non-linear dissipative hydrodynamics
of a rotating superfluid, including the elasticity of the vortex lattice.  The
basic degrees of freedom are the macroscopic velocity field ${\vec v}\,(\vec
r\,,t)$ and the macroscopic field $\vec\epsilon\,(\vec r\,,t)$ that describes
long wavelength displacements of the vortices in the transverse ($x,y$)
direction from their equilibrium locations.  (In general, the local vortex
line velocity in the transverse direction, $\dot{\vec\epsilon}\,$, does not
equal ${\vec v}_\perp$, the superfluid velocity in the plane.)  Prior to
calculating the thermal fluctuations of the lattice we summarize the equations
of motion obeyed by $\vec v$ and $\vec\epsilon$, and the long wavelength modes
of the vortex lattice.

    In order to bring out the character of the equations for the modes,
we assume, as in Ref. \cite{BC1}, that the vortex lines have an effective
mass per unit length, and thus carry a normal fluid density $\rho^*$, which
we assume to be $\ll \rho$, the mass density of the fluid \cite{mass};
in the end we take $\rho^*\to 0$.  The mass current, $\vec j$, is given by
\begin{eqnarray}
\vec j = \rho \vec v + \rho^*(\dot{\vec\epsilon} - {\vec v}_\perp);
\label{1}
\end{eqnarray}
for incompressible flow,
\begin{eqnarray}
\vec\nabla\cdot\vec j = 0.
\label{2}
\end{eqnarray}
The dynamics of the system are specified by the superfluid acceleration
equation and the law of conservation of momentum.  The linearized
superfluid acceleration equation in the rotating frame is \cite{BC1}
\begin{eqnarray}
{{\partial \vec v}\over{\partial t}} + 2\vec\Omega \,\times \dot{\vec
\epsilon} = -\vec \nabla \mu',
\label{3}
\end{eqnarray}
where $\mu' = \mu - (\vec \Omega \times \vec r\,)^2/2$ and $\mu$ the
chemical potential per unit mass.

    This equation is the time derivative of the relation between the
macroscopically averaged phase of the condensate wave function and the
macroscopically averaged flow velocity
and vortex displacements; as we show in Appendix A, for small displacements of
the vortex lattice,
\begin{eqnarray}
\vec v\, + 2\vec\Omega \,\times \vec \epsilon\, = \frac{\hbar}{m}\vec\nabla
\phi.
\label{phase}
\end{eqnarray}
Equation (\ref{phase}) generalizes the usual relation, $\vec v\, =
{\hbar/m}\vec \nabla \phi$, to the rotating system, where the flow velocity
can have non-zero vorticity as a consequence of local variations of the vortex
density.  Equation (\ref{phase}) is the analog in the rotating system of the
relation between the flow velocity and phase in a superconductor in the
presence of a vector potential,
\begin{eqnarray}
\vec v\, + \frac{e}{mc}\vec A\, = \frac{\hbar}{m}\vec\nabla
\phi.
\label{scphase}
\end{eqnarray}
Comparing with Eq.  (\ref{3}), we see that $\mu'$ is related to the
phase in the usual way,
\begin{eqnarray}
\mu'(\vec r\,,t)
= -\frac{\hbar}{m}\frac{\partial \phi(\vec r\,,t)}{\partial t}.
\label{mu-phi}
\end{eqnarray}

    The linearized equation for conservation of momentum in the rotating frame
is \cite{BC1}
\begin{eqnarray}
{{\partial \vec j}\over{\partial t}}
 + 2\vec\Omega \,\times {\vec j} + \nabla P' =
 -{\vec \sigma}_{el} - \vec\zeta,
\label{4}
\end{eqnarray}
where $P'=P-\rho(\vec \Omega \times \vec r\,)^2/2$ and $P$ is the
pressure; $-{\vec\sigma}_{el}$ is the elastic force density (directed in the
transverse plane) arising from deformations of the vortex lattice, given in
terms of the elastic energy density, $E_{el}$, of the vortex lattice by
\begin{eqnarray}
{\vec\sigma}_{el} = {{\delta E_{el}}\over {\delta \vec \epsilon}} =
{{\hbar\Omega\rho}\over{4m}}[2\vec\nabla_\perp(\vec\nabla\cdot\vec\epsilon\,)
-\nabla_\perp^2\vec\epsilon\,] -
2\Omega\lambda{{\partial^2\vec\epsilon}\over{\partial z^2}}.
\label{5}
\end{eqnarray}
Here $2\Omega\lambda = n_v E_v$ is the vortex line energy measured per
unit volume, where $E_v$ is the energy per unit length of a vortex line, and
$n_v = m\Omega/\pi\hbar$ is the number of lines per unit area.  (The detailed
elastic constant multiplying $\nabla_\perp^2\vec\epsilon\,$ in the right side
of (\ref{5}) is specific to a triangular lattice.)  We include in (\ref{4}) an
external
driving force $-\vec\zeta\,(\vec r, t)$ acting on the lattice, derived from an
external perturbation $H'=\vec\zeta\cdot\vec\epsilon$, to facilitate
calculation of the displacement autocorrelation functions.

    At low temperatures one may neglect the normal fluid mass density
associated with the bulk excitations (phonons and rotons) of the fluid.  Then
in Eq.  (\ref{4}), $\vec\nabla\, P' = \rho \vec\nabla\, \mu'$, which we
eliminate using (\ref{3}); keeping $\rho^*$ only in the inertial term we find
\begin{eqnarray}
\rho^*{{\partial}\over{\partial t}}(\dot{\vec\epsilon}-\vec v_\perp\,)
-2\rho\vec\Omega\,\times (\dot{\vec\epsilon}-\vec v_\perp) =
 -{\vec \sigma}_{el} - \vec\zeta.
\label{6}
\end{eqnarray}
To compute the superfluid velocity in terms of $\vec\epsilon$, we assume
a plane-wave spatial dependence with wave vector $\vec k$, and from (\ref{2})
and (\ref{phase}) find, for $\rho^*\ll \rho$, that
\begin{eqnarray}
\vec v +2\vec\Omega\,\times\vec\epsilon =
\hat k [\hat k\cdot(2\vec \Omega\,\times\vec\epsilon)],
\label{7}
\end{eqnarray}
where $\hat k$ is the unit vector in the $\vec k$ direction.  This equation
implies that a displacement of the vortex lattice, even uniform, induces a
flow velocity, a consequence of the lack of Galilean invariance in a rotating
system.  The quantity on the right side of (\ref{7}) is the variation of the
macroscopically averaged phase of the order parameter.

    The equations of motion are most simply written in terms of the
longitudinal and transverse displacements defined by $\epsilon_L=\hat
q\cdot\vec\epsilon$ and $\epsilon_T = \hat z\cdot(\hat
q\times\vec\epsilon\,)$, where $\hat q$ is the unit vector in the $x,y$ plane
along the direction of the projection of $\vec k$ in the plane.  (The limit of
$\vec k$ along the rotation axis presents no ambiguities.)  Substituting $\vec
v\,$ from Eq.  (\ref{7}) into (\ref{6}) and taking longitudinal and transverse
components,
we derive the coupled equations of motion for $\epsilon_L$ and $\epsilon_T$:
\begin{eqnarray}
\rho^*{\ddot \epsilon}_L +2\rho\Omega{\dot\epsilon}_T +\alpha_L\epsilon_L &=&
-\hat q\cdot\vec\zeta \equiv -\zeta_L,\nonumber\\
\rho^*{\ddot \epsilon}_T -2\rho\Omega{\dot\epsilon}_L +\alpha_T\epsilon_T &=&
-\hat z\cdot(\hat q\times\vec\zeta) \equiv -\zeta_T,
\label{8}
\end{eqnarray}
where
\begin{eqnarray}
\alpha_L(\vec k\,) &=& 4\Omega^2\rho
-{{\hbar\Omega\rho}\over{4m}}k_\perp^2+2\Omega\lambda k_z^2,\nonumber\\
\alpha_T(\vec k\,) &=& 4\Omega^2\rho {\hat k}_z^2
+{{\hbar\Omega\rho}\over{4m}}k_\perp^2+2\Omega\lambda k_z^2.
\label{99}
\end{eqnarray}
Equations (\ref{8}) agree in content with those given in Ref. \cite{BC1}.
Note that they imply that the energy
\begin{eqnarray}
E_V = \sum_{\vec k}{1\over2}[\rho^*(\dot\epsilon_L^2+ \dot\epsilon_T^2) +
\alpha_L\epsilon_L^2 + \alpha_T\epsilon_T^2]
\label{10}
\end{eqnarray}
is conserved, in the absence of external perturbations.  The fact that the
energy depends directly on the displacements, not only on their
derivatives, as is the situation in a normal lattice, is a consequence of
a vortex displacement inducing a flow velocity, Eq. (\ref{7}).

    The frequencies $\omega$ of the normal modes of the system,
determined by the four roots of the secular equation
\begin{eqnarray}
D(\vec k, \omega) \equiv (\rho^*\omega^2-\alpha_L)(\rho^*\omega^2 - \alpha_T)
- 4\Omega^2\rho^2\omega^2 = 0,
\label{11}
\end{eqnarray}
correspond to a high frequency inertial mode, of frequency given for small
$k$ by
\begin{eqnarray}
\omega_I^2 = \left(2\Omega\rho/\rho^*\right)^2,
\label{12}
\end{eqnarray}
and a generalized Tkachenko mode of frequency given for small $k$ by
\begin{eqnarray}
\omega_T^2 = {{\alpha_L\alpha_T}\over{4\Omega^2\rho^2}}=
 (2\Omega\cos\theta)^2 +
\left[{2\Omega\lambda\over\rho}\cos^2\theta(1+\cos^2\theta) +
{{\hbar\Omega}\over{4m}}\sin^4\theta\right]k^2,
\label{13}
\end{eqnarray}
where $\theta$ is the angle between the wavevector $\vec k$ and the rotation
axis. At $\theta=\pi/2$, Eq. (\ref{13}) reduces to $\omega_T^2 =
(\hbar\Omega/4m)k^2$, Tkachenko's original result.

\section{Displacement correlation functions and lattice stability}

    To calculate the effect of thermal excitation of Tkachenko modes on the
displacements of the lattice, we first construct the space-time Fourier
transform of the retarded-commutator correlation function from the relation
\begin{eqnarray}
\langle\epsilon_i\epsilon_j\rangle = \hbar{{\delta
\langle\epsilon_i\rangle}\over
{\delta \zeta_j}}.
\label{14}
\end{eqnarray}
Solution of Eqs. (\ref{8}) for $\epsilon$ in terms of $\zeta$,
\begin{eqnarray}
\left(\matrix{\epsilon_L\cr \epsilon_T\cr}\right) = {1\over D(\vec k,\omega)}
\left(\matrix{\rho^*\omega^2-\alpha_T&-2i\Omega\rho\omega\cr
      2i\Omega\rho\omega&\rho^*\omega^2-\alpha_L\cr}\right)
\left(\matrix{\zeta_L\cr \zeta_T}\right),
\label{15}
\end{eqnarray}
yields the correlation functions
\begin{eqnarray}
\langle\epsilon_L\epsilon_L\rangle(\vec k,\omega) &=& {{\hbar(\rho^*\omega^2 -
\alpha_T)}\over{D(\vec k,\omega)}},\nonumber\\
\langle\epsilon_T\epsilon_T\rangle(\vec k,\omega)&=&
 {{\hbar(\rho^*\omega^2 - \alpha_L)}\over{D(\vec k,\omega)}},\nonumber\\
\langle\epsilon_L\epsilon_T\rangle(\vec k,\omega)&=&
\langle\epsilon_T\epsilon_L\rangle(\vec k,\omega)^*
= -{{2i\hbar\Omega\rho\omega}\over{D(\vec k,\omega)}}.
\label{16}
\end{eqnarray}

    The correlation functions (\ref{16}) are conveniently written in terms of
their spectral weights, defined by
\begin{eqnarray}
\langle\epsilon_i\epsilon_j\rangle(\vec k,\omega) = \int_{-\infty}^{\infty}
{{d\omega'}\over{2\pi}}{{C_{ij}(\vec k,\omega')}\over{\omega-\omega'}}.
\label{17}
\end{eqnarray}
The weight functions obey the symmetries
$C_{LL}(\vec k,-\omega)=-C_{LL}(\vec
k,\omega)$, and $C_{TT}(\vec k,-\omega)=-C_{TT}(\vec k,\omega)$, while
$C_{LT}(\vec k,-\omega)=C_{TL}(\vec k,\omega)$; as we find from
(\ref{16}), the weights are given, for $\omega>0$, by
\begin{eqnarray}
C_{LL}(\vec k,\omega) &=&{{\pi\hbar}\over{2\Omega\rho}}\delta(\omega-\omega_I)
+
{{\pi\hbar\omega_T}\over{\alpha_L}}\delta(\omega-\omega_T),\nonumber\\
C_{TT}(\vec k,\omega) &=&{{\pi\hbar}\over{2\Omega\rho}}\delta(\omega-\omega_I)
+{{\pi\hbar\omega_T}\over{\alpha_T}}\delta(\omega-\omega_T),\nonumber\\
C_{LT}(\vec k,\omega)
& =&{{\pi\hbar}\over{2i\Omega\rho}}[\delta(\omega-\omega_I)
    -\delta(\omega-\omega_T)].
\label{18}
\end{eqnarray}

    The equal-time spatial correlations of the displacements of the vortex
lines in the rotating fluid at temperature $T = 1/\beta$ are given by
\begin{eqnarray}
\langle\epsilon_i\,(\vec r,t)\epsilon_j\,(\vec r\,',t)\rangle =
\int {{d^3k}\over{(2\pi)^3}}{{d\omega}\over{2\pi}}e^{i\vec
k\cdot\vec R}C_{ij}(\vec k,\omega)
(1+f(\omega)),
\label{19}
\end{eqnarray}
where $i,j$ denotes the transverse or longitudinal components,
$\vec R \equiv \vec r - \vec r\,'$, and
$f(\omega)= (e^{\beta\hbar\omega}-1)^{-1}$.
Because $\alpha_L$ is non-zero for $\vec k \to 0$, the longitudinal
correlation function $\langle\epsilon_L(\vec r\,)\epsilon_L(\vec
r\,')\rangle$ has a finite range of order several vortex spacings; however,
since $\alpha_T$ vanishes for in-plane wavevectors approaching zero, the
transverse correlation function $\langle\epsilon_T(\vec r\,)\epsilon_T(\vec
r\,')\rangle$ falls as $1/R_{\perp}$ in the $x,y$ plane.  Since for general
$\vec k$, $\langle\vec\epsilon\,^2\rangle=
\langle\epsilon_T^2+\epsilon_L^2\rangle$, we see that in the limit
$\rho^*\to0$, at equal times,
\begin{eqnarray}
\langle(\vec\epsilon\,(\vec r)-\vec\epsilon\,(\vec r\,'))^2\rangle
= \int {{d^3k}\over{(2\pi)^3}}
(1-\cos\vec k\cdot\vec R)\hbar\omega_T\left({1\over{\alpha_T}}+
{1\over{\alpha_L}}\right) (1+2f(\omega_T)).\nonumber\\
\label{20}
\end{eqnarray}

     Were we to take into account only two-dimensional displacements of the
vortex lattice, i.e., restrict $\vec k$ to the transverse plane, we would
find, rather,
\begin{eqnarray}
\langle(\vec\epsilon\,(\vec r)-\vec\epsilon\,(\vec r\,'))^2\rangle =
 \int {{d^2k}\over{(2\pi)^2 Z}}
(1-\cos\vec k\cdot\vec R)\hbar\omega_T\left({1\over{\alpha_T}}+
{1\over{\alpha_L}}\right) (1+2f(\omega_T)),\nonumber\\
\label{21}
\end{eqnarray}
where $Z$ is the container thickness in the $z$ direction ($Z\rho$ is the
mass per unit area).  Equation (\ref{21}) implies that Tkachenko modes in the
transverse plane should, formally, prevent formation of a stable
two-dimensional lattice.  The infrared contribution to the integral
in the limit of large separation $R =|\vec r-\vec r\,'|$ is
\begin{eqnarray}
\langle(\vec\epsilon\,(\vec r)-\vec\epsilon\,(\vec r\,'))^2\rangle
\sim {2T\over{Z\rho}} \int_0 {{d^2k}\over{(2\pi)^2}}
{{(1-\cos\vec k\cdot\vec R)}
\over{\omega_T^2}} \sim
\int_{R^{-1}} {{dk}\over k},
\label{22}
\end{eqnarray}
which diverges as $\ln R$.  The instability in two dimensions is driven by
the softness of the energy associated with shearing, $\sim\alpha_T$, for $\vec
k\,$ lying in the $x,y$ plane (see Eq. (\ref{10})), leading to a $\ln R$
divergence
in the transverse correlation function.  As in three dimensions, the
longitudinal displacement correlation function has finite range.  The mean
square displacement of a vortex from its equilibrium site similarly diverges
as $\ln{\cal R}$, where ${\cal R}$ is the container radius.

    But one should note that such an instability could, in fact, manifest
itself only in films no more than tens of atomic layers thick.  The square of
the displacement as a fraction of the area per vortex becomes, to within a
constant in the logarithm,
\begin{eqnarray}
\langle\epsilon_T\,(\vec r)^2\rangle n_v \approx 3{T\over{\Theta_D}}
{1\over{k_D Z}}\ln N_v,
\label{23}
\end{eqnarray}
where $k_D$ is the Debye wavevector, defined by $\rho/m \equiv k_D^3/6\pi^2$;
$\Theta_D \equiv \hbar^2 k_D^2/2m$, and $N_v$ is the total number of vortex
lines in the system.  For $(T/\Theta_D)\ln N_v$ of order unity, the mean
displacements can become comparable to the intervortex spacing only
for $k_DZ\raisebox{-.5ex}{$\stackrel{<}{\scriptstyle\sim}$}10^2$.

    When the full three-dimensional degrees of freedom of the vortex
excitations are taken into account, we find, instead, a finite infrared
contribution,
\begin{eqnarray}
\langle\vec\epsilon\,(\vec r\,)^2\rangle&\sim&
T\int_0 {{d^3k}\over{(2\pi)^3}}
\left({1\over\alpha_T}+{1\over\alpha_L}\right)=
T\int_0 {{d^3k}\over{(2\pi)^3}}
{{1+\cos^2\theta}\over{\rho\omega_T^2}}\nonumber\\
&\to& {T\over\rho}\int_{-1}^1
{{d\cos\theta}\over{4\pi^2}}\int_0 k^2 dk
{{1+\cos^2\theta}\over{4\Omega^2\cos^2\theta+(\hbar
k^2\Omega/4m)\sin^4\theta}}.\nonumber\\
\label{24}
\end{eqnarray}
The two terms in $(1+\cos^2\theta)$ arise respectively from transverse and
longitudinal displacements.  Inclusion of the full three-dimensional spectrum
of long-wavelength fluctuations thus stabilizes the vortex lattice.
The excursions of the vortices from their equilibrium positions are made
finite by the smaller phase space for excitations of small $k$ in three
dimensions, and the non-vanishing of both the transverse and longitudinal
coefficients, $\alpha_L$ and $\alpha_T$, in the energy (\ref{10}), as $\vec k
\to
0$ with $\cos\theta>0$.  Detailed evaluation of $\langle\vec\epsilon\,(\vec
r\,)^2\rangle$ requires calculation over all $k$, not just the long-wavelength
modes.

    What are the implications of the thermal fluctuations of the lattice
vibrations for the macroscopically averaged phase $\phi(\vec r,t)$ of the
condensate wave function?  Since the system preserves a global (gauge)
invariance allowing changes of the phase by an additive constant, only
relative phases have physical meaning.  The invariant correlations of the
order parameter are given at equal time, for Gaussianly-distributed Fourier
components of the phase, by
\begin{eqnarray}
\langle e^{i\phi(\vec r)} e^{-i\phi(\vec r\,')}\rangle
= e^{-{1\over2}\langle (\phi(\vec r)- \phi(\vec r\,'))^2\rangle}.
\label{26}
\end{eqnarray}
For plane-wave spatial dependence Eq. (\ref{7}) implies that
\begin{eqnarray}
\phi = {{im}\over{\hbar k}}2\vec\Omega\cdot(\hat k\times\vec\epsilon) =
{{2im\Omega}\over{\hbar k}}\sin\theta\,\epsilon_T;
\label{25}
\end{eqnarray}
transverse fluctuations of the lattice positions give rise to phase
fluctuations of longer range (owing to the extra power of $k$ in the
denominator).
Calculating the correlation of phase differences from Eqs.
(\ref{25}) and (\ref{16}) we have,
\begin{eqnarray}
\langle (\phi(\vec r)- \phi(\vec r\,'))^2\rangle&=&
\int {{d^3k}\over{(2\pi)^3}}{{d\omega}\over{2\pi}}
(1-\cos\vec k\cdot\vec R)
\left({{2m\Omega\sin\theta}\over{\hbar k}}\right)^2 C_{TT}(\vec
k,\omega)(1+f(\omega))\nonumber\\
&=& \int {{d^3k}\over{(2\pi)^3}}\left({{1-\cos\vec k\cdot\vec R}
\over{k^2}}\right)
{{(2m\Omega\sin\theta)^2}\over{\hbar\alpha_T}}\,
 \omega_T(1+2f(\omega_T))\nonumber\\
&\sim& \frac{\sqrt n_v}{k_D}\frac{T}{\Theta_D}\ln(\pi R^2 n_v).
\label{27}
\end{eqnarray}
Formally, (\ref{27}) diverges logarithmically as $R\to\infty$, leading to
the correlation function $\langle e^{i\phi(\vec r)} e^{-i\phi(\vec
r\,')}\rangle$ falling to zero as a power of $R$, as discussed originally by
Moore \cite{Moore1}.  Numerically, this falloff is insignificant in any finite
system, since the vortex lattice spacing $\propto 1/\sqrt{n_v}$ is always huge
compared with the interparticle spacing $\propto k_D^{-1}$, while
$\ln(\pi R^2 n_v)$ is less than $\ln N_v$, where $N_v$ is the total number
of vortices in the system.

    In general though, vanishing of the correlation function $\langle
e^{i\phi(\vec r)} e^{-i\phi(\vec r\,')}\rangle$ for large $R$ does not
indicate a loss of superfluidity.  As Josephson stressed, by analogy to a very
floppy ``slinky," such correlations do not tell one the fluctuations in the
winding of the phase angle around a closed loop.  Note also that correlations
of the gradients of the phase remain finite as $R\to\infty$.

    The existence of transverse excitations of the velocity field causes the
normal mass density, as measured by the moment of inertia, to equal to the
full mass density.  The normal mass density, $\rho_n$, is related to the
transverse current autocorrelation function by (see, e.g., \cite{scot}).
\begin{eqnarray}
\rho_n = - \lim_{k\to 0}\, \langle j_T j_T \rangle (\vec k\,,\omega=0).
\label{rho-n}
\end{eqnarray}
Taking $\rho^*=0$, whereupon $\vec j = \rho \vec v$, and calculating the
transverse component of the velocity from Eq.  (\ref{7}), we find $v_T =
-(2\Omega\,\times\vec \epsilon\,)_T = -2\Omega\epsilon_L$, and from Eqs.
(\ref{16}) and (\ref{18}) that
\begin{eqnarray}
\rho_n =
- \lim_{k\to 0}\, \rho^2 \langle v_T v_T\rangle (\vec k\,,\omega=0) =
- \lim_{k\to 0}\,  (2\Omega\rho)^2 \langle
\epsilon_L \epsilon_L \rangle (\vec k\,,\omega=0) = \rho.
\label{rho-n1}
\end{eqnarray}
Tkachenko excitations replenish the ``sum rule" (\ref{rho-n}), reflecting the
fact that in the limit of a dense vortex lattice the moment of inertia of the
container of superfluid takes on its normal fluid value.  However, this result
does not imply that the superfluid density measured dynamically, e.g., in
a second sound experiment, vanishes; rather, in the non-Galilean invariant
rotating system, the two measures become independent.


    I am grateful to A. J. Leggett and C. J. Pethick for extensive
discussions, and to E. Fradkin and R. E. Peierls for critical comments.  This
research was supported in part by National Science Foundation Grant
DMR91-22385.

\vspace{0.25truein}
\begin{Large}
\begin{bf}
\noindent{Appendix A. Macroscopically-averaged phase of the order parameter}
\end{bf}
\end{Large}

    In this Appendix we derive the relation (\ref{phase}) between the
macroscopic averages of the phase of the order parameter, the flow velocity
and the vortex displacements.  Imagine displacing a single vortex line,
initally along the $z$-axis, by a small amount $\vec\varepsilon\,(z)$ in the
transverse plane.  [We denote the local velocity, displacement and potential
by $\vec V$, $\vec \varepsilon$, and $\Phi$, to distinguish them from the
long-wavelength averaged quantities, $\vec v$, $\vec \epsilon$, and $\phi$.]
The flow velocity, $\vec V$, of the fluid about the vortex line obeys
\begin{eqnarray}
\vec \nabla\times\vec V(\vec r\,) = \frac{h}{m}\int d{\vec l}\, \delta(\vec r
- \vec r(\vec l)),
\label{curl}
\end{eqnarray}
where $l$ is the distance along the line.  For small $\varepsilon$,
$d\vec l/dz = (d\vec\varepsilon\,(z)/dz,1)$
(the final component being in the $z$-direction); thus
\begin{eqnarray}
    \vec\nabla\times\vec V(\vec r\,) = \frac{h}{m}
    \left(\frac{d\vec\varepsilon\,(z)}{dz},1\right)
    \delta(x-\varepsilon_x) \delta(y-\varepsilon_y).
\label{curl1}
\end{eqnarray}
Taking the curl of (\ref{curl1}), expanding the right side to first order in
$\varepsilon$, and assuming an incompressible fluid, $\vec\nabla\cdot\vec V =
0$, we find that the first variation of the velocity obeys
\begin{eqnarray}
\nabla^2\left(\delta \vec V-\frac{h}{m}\vec\varepsilon\times{\hat z}
\delta(x)\delta(y)\right) = \frac{h}{m}
\vec\nabla \left(\left(\varepsilon_x\frac{\partial}{\partial y}
-\varepsilon_y\frac{\partial}{\partial x}\right)\delta(x)\delta(y)\right),
\label{delsq}
\end{eqnarray}
with solution
\begin{eqnarray}
\delta\vec V(\vec r\,) = \frac{h}{m}\vec\varepsilon\times{\hat z}
\delta(x)\delta(y) +\frac{\hbar}{m} \vec\nabla \delta\Phi,
\label{vPhi}
\end{eqnarray}
where
\begin{eqnarray}
\nabla^2\delta\Phi= 2\pi
\left(\varepsilon_x\frac{\partial}{\partial y}
-\varepsilon_y\frac{\partial}{\partial x}\right)\delta(x)\delta(y).
\label{Phi}
\end{eqnarray}
Since at points away from the line, the velocity is given by $\hbar/m$ times
the gradient of the phase of the order parameter, we identify $\delta\Phi$ as
the first variation of the phase of the order parameter due to displacement
of the line.

    We now sum (\ref{vPhi}) over a lattice of vortices, and carry out a long
wavelength average.  Since the average of $\vec V$ at $\vec \varepsilon\,=0$
is the uniform rotational velocity, $\vec\Omega\,\times\vec r$, the long
wavelength average of $\delta\vec V$ is the flow velocity, $\vec v$, in the
rotating frame; similarly the long-wavelength average of $\delta\Phi$ is the
macroscopically-averaged phase, $\phi$, in the rotating frame.  Using
$hn_v/m=2\Omega$ we derive Eq.  (\ref{phase}),
\begin{eqnarray}
\vec v\, + 2\vec\Omega \,\times \vec \epsilon\, = \frac{\hbar}{m}\vec\nabla
\phi.
\nonumber
\end{eqnarray}

    Equation (\ref{Phi}) is readily integrated in terms of the Bessel function
$K_0$; for a given wavevector $q$ in the $z$-direction,
\begin{eqnarray}
\delta\Phi(\vec r\,) =
\frac{\hbar}{m}\left(\varepsilon_y x -\varepsilon_x y\right)\frac{1}{r_\perp}
\frac{\partial}{\partial r_\perp} K_0(qr_\perp),
\end{eqnarray}
where $r_\perp = (x^2+y^2)^{1/2}$.  At short distances from the line,
$qr_\perp\ll 1$, $\delta\Phi(\vec r)$ is given by $(\hbar/m) \delta\tan^{-1}
\left((y-\varepsilon_y(z))/(x-\varepsilon_x(z))\right)$, i.e., the variation
of the potential for a straight vortex line evaluated at the locally displaced
position of the line.  On the other hand, far from the line, $qr_\perp\gg 1$,
the variation $\Phi(\vec r)$ vanishes exponentially as
$\exp(-qr_\perp)/r_\perp^{1/2}$.

\newpage
\vspace{0.25truein}
\begin{Large}
\begin{bf}
\noindent{Appendix B. Commutation relations of the vortex displacements}
\end{bf}
\end{Large}

    The coefficients of $1/\omega^2$ in the high frequency limits of the
correlation functions (\ref{16}) imply, as expected, that at equal time, all
components of the displacements commute
$\langle[\epsilon_i,\epsilon_j]\rangle(\vec k\,)\equiv 0$, while the
displacements and their time rates of change obey the equal-time commutation
relations
\begin{eqnarray}
\langle[\epsilon_L,{\dot\epsilon}_L]\rangle(\vec k\,) =
\langle[\epsilon_T,{\dot\epsilon}_T]\rangle(\vec k\,) =
{\hbar\over{\rho^*}},
\label{A1}
\end{eqnarray}
and
\begin{eqnarray}
\langle[\epsilon_L,{\dot\epsilon}_T]\rangle(\vec k\,) = 0.
\label{A2}
\end{eqnarray}
On the other hand, if we assume {\it ab initio} that the inertial mass
density $\rho^*$ associated with the vortex lines vanishes, then the
commutation relations take on a quite different structure.  The
correlation functions are given by
\begin{eqnarray}
\langle\epsilon_L\epsilon_L\rangle(\vec k,\omega)&
 =& -{{\hbar\alpha_T}\over{D'(\vec k,\omega)}},\nonumber\\
\langle\epsilon_T\epsilon_T\rangle(\vec k,\omega)&
=& -{{\hbar\alpha_L}\over{D'(\vec k,\omega)}},\nonumber\\
\langle\epsilon_L\epsilon_T\rangle(\vec k,\omega)&
=& -{{2i\hbar\Omega\rho\omega}\over{D'(\vec k,\omega)}},
\label{A3}
\end{eqnarray}
where $D'(\vec k,\omega) = \alpha_L\alpha_T - 4\Omega^2\rho^2\omega^2$.  At
equal time, $\langle[\epsilon_L,{\epsilon}_L]\rangle(\vec k\,) = 0 =
\langle[\epsilon_T,{\epsilon}_T]\rangle(\vec k\,),$ as before; now, however,
the longitudinal and transverse displacements no longer commute, but rather
obey \cite{EB}
\begin{eqnarray}
\langle[\epsilon_L,{\epsilon}_T]\rangle(\vec k\,) =
{{i\hbar}\over{2\Omega\rho}}.
\label{A4}
\end{eqnarray}
The corresponding spectral weights are, for $\omega>0$,
\begin{eqnarray}
C_{LL}(\vec k,\omega) =
{{\pi\hbar\omega_T}\over{\rho\alpha_L}}\delta(\omega-\omega_T),
\hskip0.25truein
C_{TT}(\vec k,\omega) =
{{\pi\hbar\omega_T}\over{\rho\alpha_T}}\delta(\omega-\omega_T),
\label{A5}
\end{eqnarray}
and
\begin{eqnarray}
C_{LT}(\vec k,\omega) =
-{{\pi\hbar}\over{2i\Omega\rho}}\delta(\omega-\omega_T).
\label{A6}
\end{eqnarray}

    How does the non-vanishing commutation relation (\ref{A4}) develop?  By
letting $\rho^*$ go to 0, or never introducing it in the first place, one is
using an effective low frequency theory valid on frequency scales $\ll
\omega_I$.  Indeed, the exact vanishing of
$\langle[\epsilon_L,{\epsilon}_T]\rangle(\vec k\,)$ comes about through a
cancellation of the weight of the two delta functions in $C_{LT}$ in
(\ref{18}).
In the effective theory, the delta functions at $\omega_I$ are absent in the
spectral weight functions (\ref{A5}) and (\ref{A6}), and the cancellation in
$C_{LT}$ no longer takes place, leaving the effective commutation relation
(\ref{A4}).  The operators of the low frequency theory are essentially
time-averaged over several periods $2\pi/\omega_I$.

    In the limit $\rho^* \to 0$, the effective Hamiltonian of the system
becomes [cf. Eq. (\ref{10})],
\begin{eqnarray}
\bar H = \sum_{\vec k}\left({1\over 2}(\alpha_L\epsilon_L^2 +
\alpha_T\epsilon_T^2) + \zeta_L\epsilon_L + \zeta_T\epsilon_T\right),
\label{A7}
\end{eqnarray}
and the equations of motion (\ref{8}) (at $\rho^*=0$) follow directly from
(\ref{A7}) using the commutation relation (\ref{A4}).  The system is
structurally that of an ensemble of harmonic oscillators, but unlike in a
normal lattice, the energy depends on the displacements directly, rather than
only on their derivatives.  Note that the first integral in the contribution
(\ref{24}) of the long-wavelength modes to the displacements follows directly
from
(\ref{A7}).

    It is instructive to compare this situation with the analogous one of
two-dimensional motion of a particle in a magnetic field, $\vec {\cal H}$, in
the guiding center approximation \cite{JL}.  Consider $\vec {\cal H}$ in the
$z$ direction and motion only in the transverse plane.  In the gauge in which
the Hamiltonian is $H = [p_x^2 + (p_y - e{\cal H}x/c)^2]/2m$, the operator
equations of motion have as integrals,
\begin{eqnarray}
x(t)&=&{{p_y(0)}\over{m\omega_L}} +
\left(x(0)-{{p_y(0)}\over{m\omega_L}}\right)\cos\omega_Lt +
{{p_x(0)}\over{m\omega_L}}\sin\omega_Lt,\nonumber\\
y(t)&=&y(0) - \left(x(0)-{{p_y(0)}\over{m\omega_L}}\right)\sin\omega_Lt
+{{p_x(0)}\over{m\omega_L}}(\cos\omega_Lt - 1),\nonumber\\
\label{A8}
\end{eqnarray}
where $\omega_L = e{\cal H}/mc$.  These solutions preserve, as they must, the
commutation relation $[x(t),y(t)]=0$.

    In studying motions varying slowly on the scale of a Larmor period, it is
useful to average over several Larmor periods, the equivalent of averaging
over several periods of the high-frequency inertial mode in the vortex
problem.  From Eqs. (\ref{A8}) we find the averaged coordinate operators,
\begin{eqnarray}
{\bar x}(t) = {{p_y(0)}\over{m\omega_L}}, \hskip0.25truein
{\bar y}(t) = y(0) - {{p_x(0)}\over{m\omega_L}}.
\label{A9}
\end{eqnarray}
These averaged operators do not, in fact, commute, but rather obey
\begin{eqnarray}
[{\bar x}(t), {\bar y}(t)] = -{i\hbar\over{m\omega_L}},
\label{A10}
\end{eqnarray}
the analog of the commutator (\ref{A4}) of the displacement operators in the
low-frequency effective theory \cite{Dirac}.

    In the presence of a driving potential $V(x,y)$, the equation of motion is
$m\ddot {\vec r} = (e/c)(\dot {\vec r}\,\times \vec {\cal H})- \nabla V$.
For slowly varying phenomena, one can similarly neglect the inertia term
here and replace the operators by their averages over several Larmor periods,
in which case the equation of motion reduces, in the limit $m \to 0$, to the
requirement that the net force on the particle vanish:
\begin{eqnarray}
{e\over c}(\dot {\vec {\bar r}}\,\times \vec {\cal H}) -\nabla
V({\bar x(t)},{\bar y(t)}) = 0,
\label{A11}
\end{eqnarray}
the analog of Eqs.  (\ref{8}) in the limit $\rho^*\to 0$.  This equation can
be derived directly from the effective Hamiltonian, ${\bar H} = V({\bar
x(t)},{\bar y(t)})$, using the commutation relation (\ref{A10}).

\bibliographystyle{unsrt}

\bibliographystyle{unsrt}

\begin{thebibliography}{99}
\bibitem{Moore1} Equation (\ref{27}) agrees qualitatively with the result
of Moore \cite{Moore} who studied effects of phase fluctuations arising from
lattice oscillations in Type II superconductors and neutral superfluids; in
charged superconductors he finds a critical dimension for destruction of phase
coherence of four, and for a neutral superfluid a critical dimension of three.
Inclusion here of the full structure of the Tkachenko modes in evaluating the
fluctuations does not modify the infrared behavior found by Moore, who assumed
an effective energy dependent only on gradients of the lattice displacements.
\end{thebibliography}

\end{document}